# Zeeman relaxation of MnH ($X^7\Sigma^+$) in collisions with $^3$He: mechanism and comparison with experiment

F. Turpin, T. Stoecklin[1] and Ph. Halvick

*Institut des Sciences Moléculaires, CNRS-UMR 5255, Université de Bordeaux,
33405 Talence, France*

**Abstract**

We present a theoretical study of the Zeeman relaxation of the magnetically trappable lowest field seeking state of MnH($^7\Sigma$) in collisions with $^3$He. We analyze the collisional Zeeman transition mechanism as a function of the final diatomic state and its variation as a function of an applied magnetic field. We show that as a result of this mechanism the levels with $\Delta M_j > 2$ give negligible contributions to the Zeemam relaxation cross section. We also compare our results to the experimental cross sections obtained from the buffer gas cooling and magnetic trapping of this molecule and investigate the dependence of the Zeeman relaxation cross section on the accuracy of the three body interaction at ultralow energies.

---

[1] Corresponding author : t.stoecklin@ism.u-bordeaux1.fr

# 1. Introduction

The recent advances of cooling and trapping experimental methods of molecules [1,2,3,4] have stimulated a wide range of theoretical studies dedicated to collisions in ultra cold molecular gases [5,6,7,8]. These ultra cold molecule samples have potential applications in many different fields like precision spectroscopic measurements [9,10,11] or quantum information storage and processing [12,13]. Photoassociation spectroscopy, magnetic tuning of Feshbach resonances and Stark deceleration [14,15,16,17] are currently applied to cool down a variety of molecular systems. However, $^{3}$He buffer gas cooling remains the most universal technique used to cool down molecules. It was recently shown, when followed by evaporative cooling, to even allow the production of a Bose Einstein condensate [18]. Ultra cold paramagnetic molecules can then be confined using a magnetic field. These two techniques were used in the first experiment in 1998 which produced large samples of cold neutral molecules (CaH [19]). The buffer gas cooling efficiency depends critically on the ratio of elastic to inelastic collision rates while the inelastic processes leading to trap loss were shown by Krems *et al.* to be produced by the spin-rotation interactions [20] for $^{2}\Sigma$ molecules and are assumed to be due to the spin-spin interaction for $\Sigma$ states of higher spin multiplicity. The theoretical studies dedicated to Zeeman relaxation of diatomic molecules in a $\Sigma$ state in collisions with $^{3}$He and submitted to a magnetic field are He-O$_2$($^{3}\Sigma$) [21], He+CaH($^{2}\Sigma^{+}$) and Ar+NH($^{3}\Sigma^{-}$) [22], He-NH($^{3}\Sigma^{-}$) [23,24,25,26] and He-N$_2^{+}$($^{2}\Sigma^{+}$)[27,28,29]. In a recent study, Bakker *et al.* [30] suggested that MnH(X$^{7}\Sigma^{+}$) is a good candidate to be cooled using buffer gas cooling and stored using magnetic trapping as this molecule has a large magnetic moment, relatively small spin-spin and spin-rotation coupling constants and a large rotational constant. Shortly after, they performed successfully a buffer-gas loading and magnetic trapping experiment of this molecule [31]. They gave an estimate of the elastic collision cross section with $^{3}$He and the ratio of the elastic to inelastic cross section. In the present calculations which aim to reproduce these results, we also investigate the specificity of the Zeeman relaxation mechanism for collisions involving a diatomic molecule in a $^{7}\Sigma$ state and determine which diatomic constant is critical in the cooling and trapping process. In section 2 we briefly recall the main steps of our Close Coupling calculations and discuss our results in section 3.

# 2. Calculations

Very recently [32], we developed an analytical model of the potential energy surface (PES) for the He-MnH collision and used it to obtain the binding energies of the $^3$He-MnH and $^4$He-MnH van der Waals complexes from Close Coupling calculations. In the present work, we use this PES model to study the collisional Zeeman relaxation of the magnetically trappable lowest field seeking state $M_J$=3 of MnH($X^7\Sigma^+$) belonging to the septuplet associated with $N$=0, $J$=$S$=3, where $N$ and $S$ and $J$ designate the quantum numbers associated with the rotational the electronic spin and the total angular momenta of MnH excluding nuclear spin while $M_N$, $M_S$ and $M_J$ are the quantum numbers associated with their projections along the $Z$ space fixed axis.

The Close Coupling equations which we solve are a simple extension of the method developed by Krems and Dalgarno [22] to treat the collisional spin depolarisation of a diatomic molecule in a $^3\Sigma$ state in collision with a structureless atom. Here we use the rigid rotor approximation so the Hamiltonian can be written like:

$$H = -\frac{\hbar^2}{2\mu_{A-BC}}\left(\frac{1}{R}\frac{d^2}{dR^2}R\right) + \frac{\vec{L}^2}{2\mu_{A-BC}R^2} + H_{Diatom} + V(R,\theta) \quad (1)$$

Where $\vec{L}^2$ is the angular momentum associated with the intermolecular coordinate $\vec{R}$ and $H_{Diatom}$ is the hamiltonian for the MnH($^7\Sigma$) monomer. We neglect the hyperfine structure of MnH and all the terms of the diatomic effective Hamiltonian given by Gengler *et al.* [33] the magnitudes of which are smaller than $10^{-4}$ cm$^{-1}$.
We write

$$H_{Diatom} = B\vec{N}^2 - D\vec{N}^4 + \gamma \vec{N}\cdot\vec{S} + \frac{2}{3}\lambda_{SS}\sqrt{\frac{24\pi}{5}}\sum_q Y_2^{q*}\left[\vec{S}\otimes\vec{S}\right]_q^2 + g\mu_0 \vec{B}\cdot\vec{S} \quad (2)$$

where B and D are respectively the rotational constant and the first centrifugal distortion constant while $\gamma$, $\lambda_{SS}$ and g are the spin-rotation, the spin-spin interaction constants and the g factor of the diatomic molecule the values of which are given in reference [33]. $\mu_0$ and $\vec{B}$ are the Bohr magneton and the magnetic field defining the Z space fixed axis. The neglect of the hyperfine structure of MnH is all the more so justified since the rotational constant of MnH is large (B= 5.606) whereas the spin-rotation constant (3.073 $10^{-2}$ cm$^{-1}$) which is small is still three times larger than the largest hyperfine constant (b$_F$(Mn)=9.284 $10^{-3}$ cm$^{-1}$). Furthermore Steimle et al [34] showed (see Fig. 5 of their paper dedicated to the Zeeman effect for this molecule) that the nuclear spin decouples from the molecular axis for a magnetic field larger than 1700 Gauss making the hyperfine effects negligible. Also, the neglect of the hyperfine structure of MnH seems to be appropriate to model the experiment of Bakker *et al*. [30] as the value of magnetic field which they use is 20000 Gauss.

The diagonalisation of the diatomic Hamiltonian (2) $\left[ CH_{Diatom} C^{-1} \right]_{\alpha\beta} = \xi_\alpha \delta_{\alpha\beta}$ in an uncoupled basis set $\phi_i = \chi_{v,N}(r) | NM_N \rangle | SM_S \rangle$ gives the diatomic energies represented in Fig.1 as a function of the magnitude of the applied magnetic field.

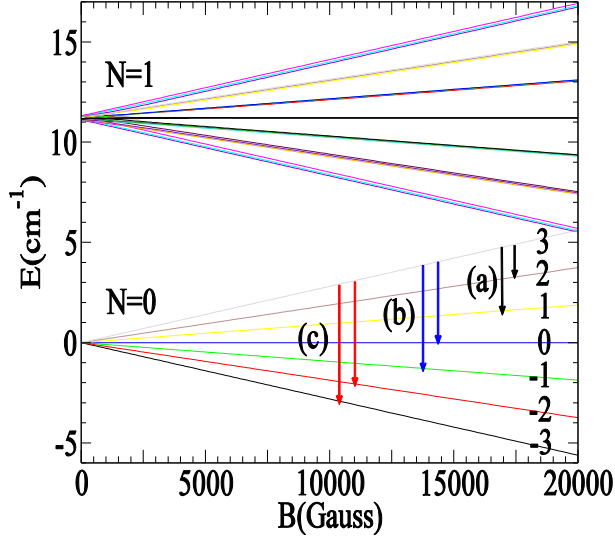

*FIG. 1 (color online)*. Diatomic eigen energies of the Hamiltonian (2) as a function of the applied magnetic field. The value of $M_J$ is reported on each curves of the N=0 septuplet. The Zeeman relaxation transitions originating from the (N=0, $M_J$ =3) level are also represented.

Each energy level $\xi_\alpha$ of the diatomic molecule is associated with a single value of $M_J = M_N + M_S$ denoted $M_J(\alpha)$. For a given value $M_T$ of the projection of the total angular momentum along the direction of the *z* space fixed axis and for a given $M_J(\alpha)$, the projection of the relative angular momentum *L* along the *z* space fixed axis is then simply $M_L = M_T - M_J(\alpha)$. The basis set describing the collision process is then obtained by adding the possible values of the quantum number *L* for each value of $\alpha$. This basis set is denoted by the quantum numbers $\alpha$, $M_L$, and *L*. In this basis set, the close coupling equations which have to be solved take the form

$$\left[ \frac{d^2}{dR^2} - \frac{L(L+1)}{R^2} + 2\mu [E - \xi_\alpha] \right] F_{\alpha, M_L(\alpha), L}(R) = 2\mu \sum_{\alpha', M_L'(\alpha'), L'} \left[ C^T UC \right]_{\alpha, M_L(\alpha), L}^{\alpha', M_L'(\alpha'), L'} F_{\alpha', M_L'(\alpha'), L'}(R) \quad (3)$$

demonstrated in reference [22]. The parameters used to perform these calculations can be found in our recent work dedicated to the He-MnH bound states [32]. As we are interested in the very low collision energy range, the Zeeman relaxation cross sections are calculated for the basis set associated with the single value $M_T = M_J$ of the Z space fixed projection of the total angular momentum and for the positive parity of the system which includes the *s* wave. The Zeeman cross sections were verified to converge to better than 1% as a function of the size of the rotational and relative basis sets and of the radial propagation parameters.

## 3. Results
### 3.1 Zeeman relaxation mechanism

We first consider the specific details of the Zeeman transition mechanism in the case of a $^7\Sigma$ molecule. At zero field, the seven lowest rotational energy levels associated with ($N=0$, $J=S=3$, $-3 \leq M_J \leq 3$) are degenerate but we will see that the mechanism of Zeeman relaxation from the lowest field seeking state level $M_J = 3$ is not the same for all the final levels. The corresponding transitions under study are represented in Fig.1. Each rotational level is associated with a given value of $M_J = M_N + M_S$ which can be obtained for different combinations of the possible values of $M_N$ and $M_S$. As the spin-spin operator couples $N$ with $N \pm 2$, the fundamental rotational levels have $N=0$ and $N=2$ components. The main components of each of the seven fundamental rotational levels correspond to $N=M_N=0$ which gives $M_S=M_J$, but they have also minor components associated with $N=2$. For example the main components of the level $M_J = 3$ is associated with $N=M_N=0$, $M_S=M_J=3$ but it has also three $N=2$ minor components associated with ($M_N=0$, $M_S=3$), ($M_N=1$, $M_S=2$) and ($M_N=2$, $M_S=1$) as illustrated in Table 1 where the eigenvector coefficients associated with the $M_J=3$ diatomic state in the $|NM_N\rangle|SM_S\rangle$ basis set are reported.

The intermolecular potential couples only levels with the same value of $M_S$. Furthermore, the main components of two different levels cannot be coupled and the coupled levels must comply with the rule:

$$\Delta M_J = M_{J'} - M_J = \Delta M_N = M_{N'} - M_N \qquad (4)$$

We can then first distinguish a group of final levels ($M_{J'}$) which have a minor component (along $N'=2$, $M_{N'}$ with $-2 \leq M_{N'} \leq 2$ ) with the same value of $M_{S'} = 3$ as the main component of the initial level ($M_J = M_S = 3$, $N = M_N = 0$). As a result, applying rule (4) means that $M_{N'} + 3 = M_{J'}$. These are the two levels $M_{J'} = M_J - 1 = 2$ ($M_{N'} = -1$) and $M_{J'} = M_J - 2 = 1$ ($M_{N'} = -2$) which we call levels of group (a). The Zeeman transitions to these levels then follow an indirect mechanism due to the spin-spin interaction as they are mediated by the rotational level $N'=2$ and the corresponding Zeeman transitions cross sections decrease in magnitude when $\Delta M_J$ increases. We checked this analysis which stresses the central role played by the spin-spin interaction by changing the value of the spin-spin constant. We found that the Zeeman transition cross sections towards these levels indeed vary like the square of the spin-spin constant. Moreover, these cross sections change very little when the spin-rotation constant is varied. The four remaining levels cannot have minor components corresponding to $M_{S'} = M_S = 3$ and can only be coupled with the minor components of the initial level ($M_J =3$). As a consequence, the magnitudes of the corresponding transition cross sections to these levels are smaller than those of group (a) by more than ten orders of magnitude as can be seen in Fig.1. This simple analysis shows that collisional Zeeman relaxation for a $^7\Sigma$ molecule is dominated by the transitions $\Delta M_J = M_J - M_{J'} =1$ and 2 which are those of group (a). The same result was found by Stoll [35] from the numerical simulation of the MnH buffer gas cooling experiment. Among the four remaining final levels only two have minor components with the same value of $M_{S'}$ (in bold) as the minor components of the initial level. They are the levels $M_{J'} =M_J - 3=0$ ($M_{S'}=$-2,-1,0,**1,2**) and $M_{J'} =M_J - 4= -1$ ($M_{S'}=$-3,-2,-1,0,**1**) which we call levels from group (b). These levels are also coupled together as they have minor components in common. The two remaining levels $M_{J'} =M_J - 5 = -2$ ($M_{S'}=$-3,-2,-1,0) and $M_{J'} =M_J - 6 = -3$ ($M_{S'}=$-3,-2,-1) which we call levels from the group (c) do not have any minor components in common with the initial level. As a result the matrix elements of the potential between the levels of group (c) and the initial level are zero. They are however coupled together and with the levels of group (b) as they have minor components in common. The Zeeman transitions to the levels of group (c) are then mediated by those of group (b) and as such are strongly correlated. We checked again this analysis by changing the value of the spin-spin constant and found indeed that the Zeeman transition cross sections towards the levels of group (b) and (c) vary with the spin-spin constant at the power four and six respectively. This unusual coupling mode gives rise to the interference patterns at very low collision energy which can be seen in Fig.2.

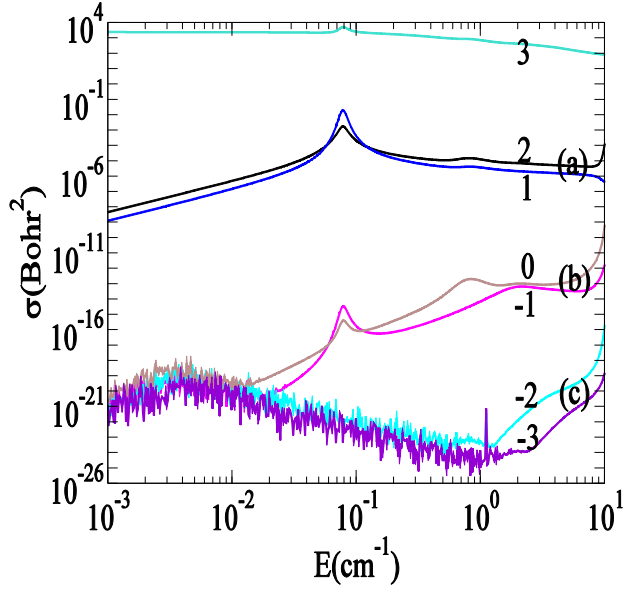

**FIG. 2** *(color online). Cross sections for the Zeeman transitions in MnH(N=0, $M_j=3$)-$^3$He collisions as a function of collision energy. The final value of $M_J$ is reported on each curve as well as the group of transitions (a),(b),(c) defined in the paragraph dedicated to the Zeeman relaxation mechanism*

Because of this indirect coupling mechanism between the initial level and the levels of group (c), the Zeeman transition cross section to the levels of group (c) are also smaller by several orders of magnitude than those of group (b) when the collision energy becomes larger than the coupling between the levels of groups (b) and (c). When a magnetic field is applied the rotational energy levels are split and the three groups become well separated, even in the limit of very low collision energy as can be seen in Figs. 3, 4 and 5 respectively for B= 1, 1000 and 20000 Gauss.

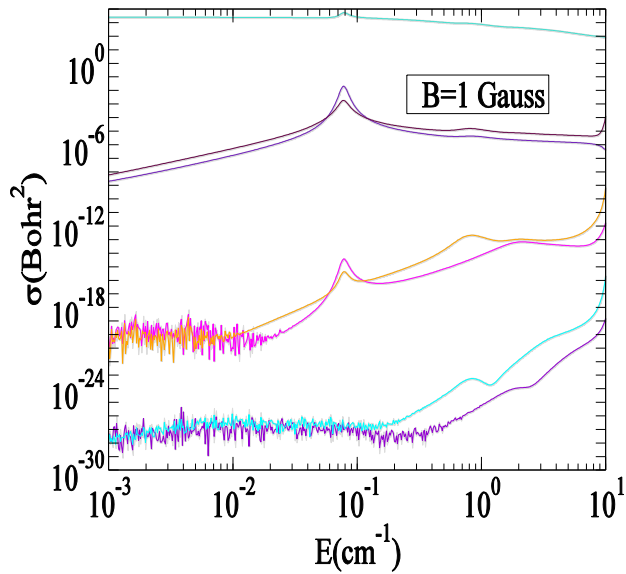

FIG. 3 (color online). Cross sections as a function of collision energy for the Zeeman transitions in MnH($N=0$, $M_j=3$)-$^3$He collisions subjected to a magnetic field of 1 Gauss.

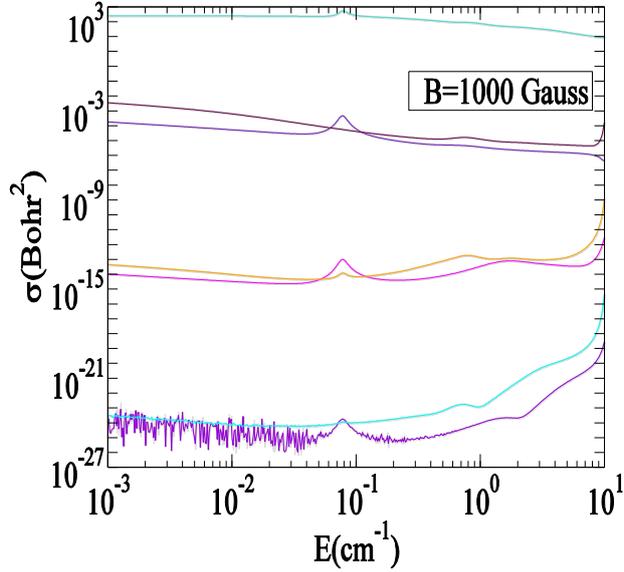

FIG. 4 (color online). Cross sections as a function of collision energy for the Zeeman transitions in MnH($N=0$, $M_j=3$)-$^3$He collisions subjected to a magnetic field of 1000 Gauss.

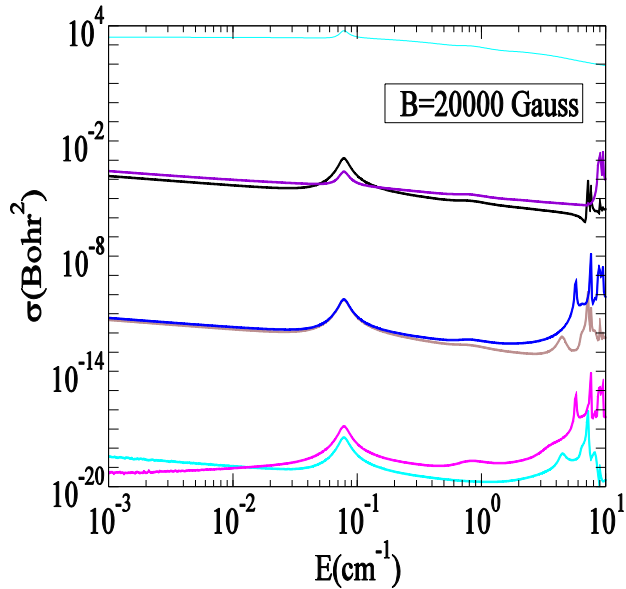

FIG. 5 (color online). Cross sections as a function of collision energy for the Zeeman transitions in MnH($N=0$, $M_j=3$)-$^3$He collisions subjected to a magnetic field of 20000 Gauss.

The interference patterns vanish progressively when the magnitude of the field is increased as a result of the increase of the splitting of the interacting levels. This progressive disappearance shows that the oscillations of the cross sections are not simple numerical artefacts but result from the coupling between the levels of groups (b) and (c). In the first part of the present analysis we concluded that Zeeman relaxation for a $^7\Sigma$ molecule is dominated by the transitions $\Delta M_J = M_J - M_{J'} = 1$ and 2 which are those of group (a). This is seen in Fig. 6 where the two total parity components (the parity is equal to $(-1)^{N+1}$) of the Zeeman relaxation and elastic cross sections for the $M_J = 3$ state of MnH are represented at zero field.

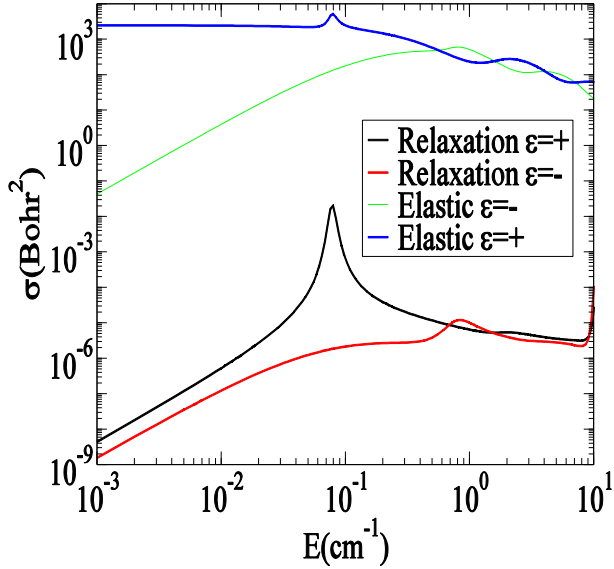

*FIG. 6 (color online). Cross sections for elastic and Zeeman relaxation in MnH(N=0, $M_J$ =3)-$^3$He collisions as a function of collision energy. The two parity ($\varepsilon$) components are represented.*

The interference patterns which involve the transitions to the levels of groups (b) and (c) are completely wiped out as the contributions of these levels to the global Zeeman relaxation cross section are negligible. This figure also shows two main resonances which are both shape resonances. The first resonance appears on the positive parity curve and is then associated with $l=2$ while the second appears on the negative parity curve and is then associated with $l=3$. The ratio of the elastic to the inelastic cross section which is the limiting factor for the cooling and trapping process clearly reaches its maximum for the main resonance around E=0.1 cm$^{-1}$. This aspect will be discussed in the next paragraph dedicated to the comparison with the experimental data. In Fig. 7 where we reported the Zeeman relaxation cross section as a function of the applied magnetic field, we notice the usual progressive removal of the Wigner's law suppression as the field increases in strength in the [10$^{-3}$, 10$^{-1}$] cm$^{-1}$ energy range. The explanation of this mechanism which was proposed by Tiesinga *et al* [36] and Volpi and Bohn [37] is based on the removal of the degeneracy of the initial and final channels when the field is applied. The elastic cross sections which are represented in Fig.2 are not reported here as they are not modified by the action of the magnetic field. We can also notice new resonances above 3 cm$^{-1}$ on the curve associated with the highest value of the applied magnetic field (B=20000 Gauss) considered in this paper. They result from the crossing with the field dressed $N$=1 diatomic states which occur around the value of the field used in the experiment of Meijer and co-workers as illustrated in Fig.1. We can also see that the magnitude of the main resonance decreases when the applied magnetic field increases. An interesting feature can also be noticed for B=100 Gauss as the main resonance appears to be shifted and broadened. As we know from the previous analysis, the two main contributions to the Zeeman cross section are the transitions $\Delta M_J = M_J - M_{J'} = 1$ and 2. As can be seen in Fig. 8 for this value of the field, the transition cross section towards the level $M_J$ =2 is larger than the one leading to the level $M_J$ =1 and is structured with a first peak at lower collision energy. This peak can be interpreted as a Feshbach resonance between the levels $M_J$=3 and $M_J$=1 which is superimposed on the shape resonance.

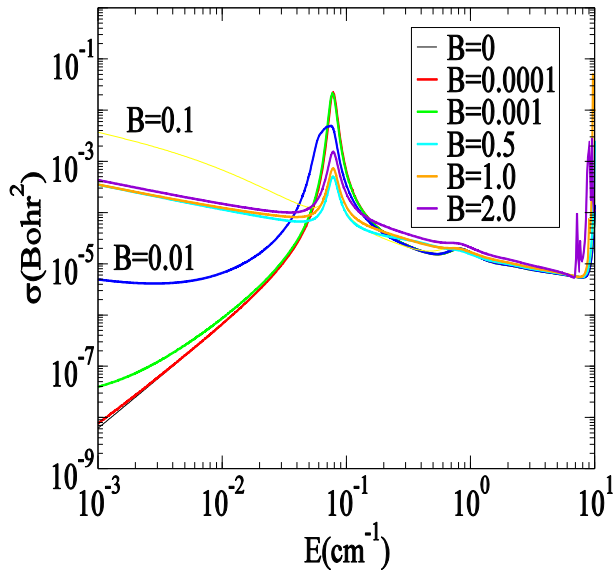

*FIG. 7 (color online). Cross sections for Zeeman relaxation in MnH(N=0, $M_j$=3)-$^3$He collisions as a function of collision energy and the applied magnetic field in Gauss.*

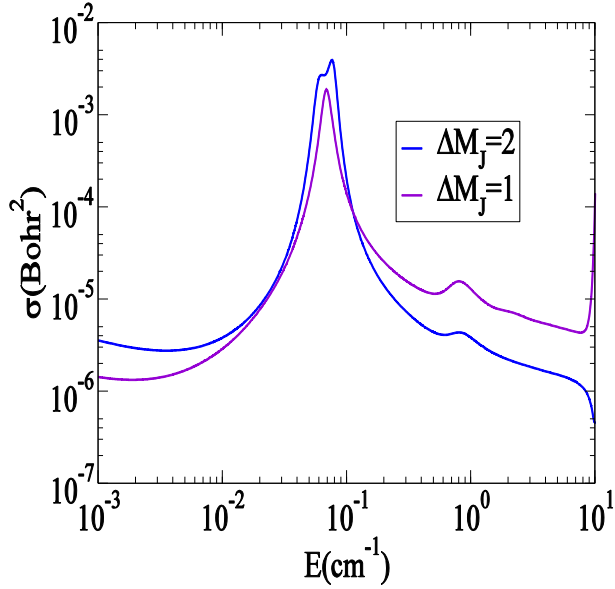

*FIG. 8 (color online). Cross sections as a function of collision energy for the Zeeman transitions $\Delta M_J = M_J - M_{J'} = 1$ and 2 in MnH($N=0$, $M_J = 3$)-$^3$He collisions subjected to a magnetic field of 100 Gauss.*

### 3.2 Comparison with experiment:

The estimated experimental temperature is 650 mK. If we take a collision energy E=0.45 cm$^{-1}$ we find a value of $\sigma$=314 Å$^2$ for the elastic cross section at zero field in the middle of the trap. This value compares well with the experimental evaluation: $90 \leq \sigma_{Exp} \leq 190$ Å$^2$. For the same values of the collision energy, we also calculate the ratio of the elastic to inelastic cross section which has to be large in order to allow the cooling and trapping processes. The calculated value for this energy is 1000000 whereas its experimentally evaluated value is only 500. This experimental value was obtained by M. Stoll [35] from trajectory simulations of MnH in a buffer gas experiment. They compared the experimental lifetime to the time dependence of the density of molecules in the center of the trapping region which they obtained from their simulation using a step model for the inelastic cross section. The use of a more realistic model of the inelastic cross section in their simulation could certainly change the estimated experimental ratio but certainly not enough to explain such a discrepancy between the theoretical and experimental values. On the theoretical side, the limiting factor of the cooling process is known to be the minimum value of this ratio which we have seen in the previous paragraph is obtained for the energy of the shape resonance around E=0.1 cm$^{-1}$. For this energy which contributes significantly to the experimental Boltzman distribution of collision energies at 650 mK, the value of the calculated ratio is 10000 which is closer to the experimental estimate but still 20 times too large. As we estimate the global accuracy of the *ab initio* potential to be 5%, we performed test calculations using the three body interaction potential multiplied by a factor ranging from 0.95 to 1.05 in order to check the dependence of our Close Coupling results on the accuracy of the potential energy surface model. Such sensitivity tests [20,38] which have been done for other systems showed that the scattering results can be strongly modified at very low energy.

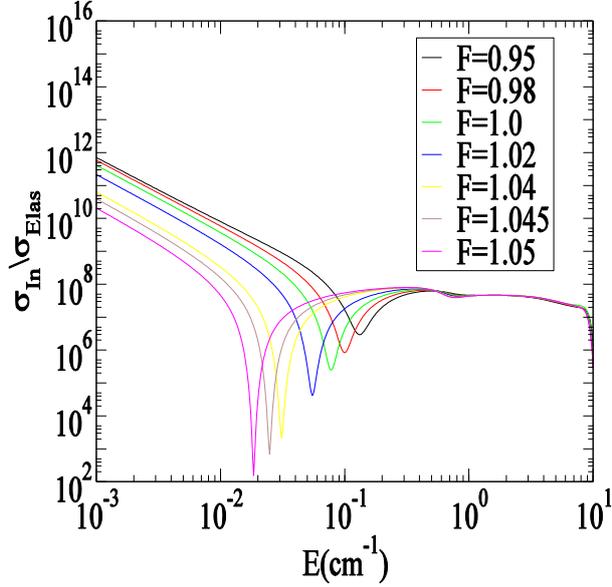

**FIG. 9** *(color online). Ratio of the elastic to inelastic cross sections of MnH(N=0, $M_j$=3) in collisions with $^3$He as a function of collision energy using our model PES multiplied by a factor F ranging from 0.95 to 1.05.*

In Fig. 9 we report the calculated elastic to inelastic cross section ratio. Here, only the very low collision energy and the resonance regions are strongly modified while the energy region above the resonance is almost unchanged. The amplitude of the resonance is found to be a monotonous function of the multiplying factor of the potential and a very good agreement with the experimental ratio can be obtained for a multiplying factor of 1.045. We will not apply this multiplicative factor to our PES for the moment however as we have seen that there is some uncertainty in the experimental evaluation of this ratio. Whatever is the exact value of the multiplicative factor which should be used, this comparison suggests that the limiting factor of the cooling and trapping experiment of MnH is a very low energy resonance.

**4. Conclusion**

We showed that the critical diatomic parameters for the buffer gas cooling and magnetic trapping of a $^7\Sigma$ molecules are the spin-spin and rotational constants as collisional Zeeman relaxation was found to be controlled by the spin-spin operator. The collisional

Zeeman relaxation is seen to be dominated by the transitions $\Delta M_J = M_J - M_{J'} = 1$ and 2 in agreement with experiment. The calculated elastic cross section is in good agreement with its experimental value while the maximum value of the calculated ratio of the elastic to inelastic cross section is found to be 20 times smaller than its experimental estimate. This last result however was showed to be strongly dependent on the accuracy of the potential energy surface as a global multiplying factor of 1.045 allowed excellent agreement between the theoretical and experimental values of this ratio. The magnitude of this factor seems to be realistic given that the current estimated accuracy of the potential energy surface model is around 5%. On the experimental side the use of a more realistic inelastic cross section than a step function may also change its estimated value from the simulation of the buffer gas cooling experiment. This comparison suggests that the limiting factor of the cooling and trapping experiment of MnH is a very low energy resonance. The possible existence of other loss channels not included in the present theory will be considered in a future work by including the hyperfine structure. Possible improvements in buffer gaz cooling of MnH will also be pursued in future calculations by considering the action of combined electric and magnetic field on this resonance.

**Table 1**: Eigenvector coefficients associated with the MnH(N=0,$M_J$=3) diatomic state in the $|NM_N\rangle|SM_S\rangle$ basis set.

| N | 0 | 2 | 2 | 2 |
|---|---|---|---|---|
| $M_N$ | 0 | 2 | 1 | 0 |
| $M_S$ | 3 | 1 | 2 | 3 |
| c | 0.999999649 | 3.42 $10^{-4}$ | -5.4 $10^{-4}$ | 5.4 $10^{-4}$ |